\documentstyle[aps,preprint]{revtex}
\begin{document}
\draft
\title{Black hole mass spectrum vs spectrum of Hawking radiation}
\author{V.A.Berezin}
\address{Institute for Nuclear Research of the Russian Academy of Sciences,
60th October Anniversary Prospect, 7a, 117312, Moscow, Russia
e-mail: berezin@ms2.inr.ac.ru}
\author{A.M.Boyarsky, A.Yu.Neronov}
\address{Dept. of Mathematics and Mechanics, Lomonosov Moscow State University,
119899, Moscow, Russia
e-mail: boyarsk@mech.math.msu.su, aneronov@mech.math.msu.su}
\date{\today}
\maketitle
\begin{abstract}
We consider a massive selfgravitating shell as a model for collapsing body 
and a null selfgravitating shell as a model for quanta of Hawking radiation.
It is show that the mass-energy spectra for the body and the radiation
do not match. The way out of this difficulty is to consider not only
out-going radiation but also the ingoing one. It means that the structure
of black hole is changing during its evaporation resulting in the 
Bekenstein-Mukhanov spectrum for large masses.
\end{abstract}

%\twocolumn

It is well known that in classical physics any spherically 
symmetric neutral black hole 
state is described by the single parameter, the total mass (Schwarzschild 
mass, or total energy) of this black hole. And after the black hole formation 
all the information about the details of the collapse is lost. In 
particular, the black hole can be formed either due to the bound motion of 
matter or due to its unbound motion. In quantum mechanics we are used to the 
fact that we have a discrete mass (energy) spectrum for a bound motion and 
a continuous spectrum for an unbound motion. We have now two possibilities. 
Either the black hole total mass has a discrete spectrum or its spectrum is 
continuous. If it is continuous, then it is not clear, first, why in the 
case of bound motion (which is, of course, also possible) the spectrum is 
continuous and, second, how to construct a ground state with nonzero mass 
which, we have reasons to believe, exists. If it is discrete, then the 
corresponding discrete quantum number has the same origin both for bound 
and unbound motions. But the bound motion has also a ``conventional'' 
quantum number (or numbers). And the unbound motion has also a 
``conventional'' continuous parameter (or parameters). From this it follows 
that if the mass of the black hole is quantized (e.i., discrete), the quantum 
black hole state is described not only by its mass but also by some other 
parameter(s), discrete or continuous.

But if this is the case, the spectrum of radiation coming from a black hole 
should have a very complicated structure. Indeed, according to the Bohr 
postulate, the spectrum of emitted quanta must correspond to the energy level 
spacing of quantum system. If the black hole spectrum depends on extra 
parameters, the radiation
spectrum would have a fine structure in the case of discrete additional 
parameters or
it is continuous in the case of continuous additional parameters. So we have
to conclude that the radiation spectrum ``remembers'' the way how the 
black  hole was formed. The spectrum seems to be not the universal one. But in
accordance with the well known semiclassical result by  S.Hawking, the spectrum of 
radiation from a black hole is universal -- it is thermal
with the temperature
$T_{BH}=m_{pl}^2/ 8\pi m$
where $m$ is  the mass of the Schwarzschild black hole. 

We are led to a contradiction. If we consider radiation propagating on the
black hole background we have to conclude that it has the same spectrum
no matter how the black hole was formed. This property mimics the classical
property that the only parameter describing a black hole is its mass. On
the other side if we calculate the spectrum of radiation from the black
hole mass spectrum, we see that it depends on how the black hole 
was formed and is different for different black holes with equal masses.

In this paper we demonstrate on a simple quantum black hole model (see 
\cite{prd57} for the detailed description) how this
contradiction could be resolved. The black hole spectrum has a complicated 
structure with all the properties discussed above. Nevertheless, the 
spectrum of radiation is universal for the black holes with Schwarzschild 
mass $m$.

The main effect which enables us to resolve the contradiction is the following.
We take fully into account the effect of back reaction of radiation on
the gravitational field of the black hole. We show that only radiation with
particular (discrete) spectrum could propagate from the black hole 
self-consistently. This turns out to be the property which defines the universal
spectrum of Hawking radiation from black hole.

We consider the following quantum black hole model. The black hole is formed 
by a  spherically symmetric collapsing 
thin shell (Fig. \ref{fig:spacetime}). There 
are different types of space-times formed either by bound (a) or unbound (b)
motion of the shell.
and
The corresponding Hamiltonian formalism has been extensively studied 
\cite{kuchar,prd57}
The result is that we have the only nontrivial equation
\cite{prd57}
\begin{equation}
 \label{exponent}
e^{ G P_R/R}
=
\frac{1}{2\sqrt{F_{in}F_{out}}}\left( F_{in}+F_{out}-
\frac
{\displaystyle M^2G^2}
{\displaystyle R^2}
\pm
\frac
{\displaystyle 2MG}
{\displaystyle R}
{\cal Z}
\right)
 \end{equation}
 \begin{equation}
 \label{z}
{\cal Z}=\sqrt{\left( F_{out}-F_{in}\right)^2\frac{\displaystyle R^2}{
\displaystyle 4 M^2G^2}-\frac{1}{2}\left( F_{out}+F_{in}\right)
+\frac{\displaystyle M^2G^2}{\displaystyle 4 R^2}}
 \end{equation}
where  $M$ is bare mass of the shell, $R$ is the radius of the shell, 
$\hat P_R$ is the corresponding 
conjugate momentum, $F_{in,out}=1-2Gm_{in,out}/ R$ and $m_{in,out}$ are 
the Schwarzschild masses inside and outside the shell. This equation
determines the trajectory of the shell $P_R (R)$ 
in the phase space of the 
model. 
Here the sign $+$ or $-$ is chosen in the right hand side of the equation
depending on whether the shell is in-going or out-going.

For the bound motion we must use the symmetric 
($P_R\rightarrow -P_R$) form of constraint.
  \begin{equation}
\label{Cons} 
F_{in}+F_{out}-
G^2 M^2 R^{-2}-\sqrt{F_{in}F_{out}}\mbox{ ch}\left(
G P_R/R\right) =0
 \end{equation}

The system which describes spherically symmetric gravitational field and a 
thin shell could be quantized in accordance with standard Dirac procedure.
Then the classical constraint (\ref{Cons}) is converted into a finite 
difference equation
 \begin{eqnarray}
 \label{main}
 \Psi (m, \mu, S+i\zeta )+\Psi (m, \mu, S-i\zeta )=\nonumber\\
\left( F_{in}
F_{out}\right)^{-1/2}
\left( F_{in}+F_{out}-M^2/4m^2S\right)
\Psi (m, \mu, S)
 \end{eqnarray} 
where $m=m_{out}$, $\mu =m_{in}/m_{out}$ and 
$S =  R^2/ (2 G m)^2
=R^2/R_g^2$ is a dimensionless variable which measure 
the surface area of the shell. The finite shift parameter 
$\zeta =m^2_{pl}/2m^2_{out}$.

The quantum system described by the equation (\ref{main}) possesses a 
discrete mass 
spectrum for the values of parameters $m_{out}<M$. The spectrum
in the large black hole approximation is
 \begin{equation}
\label{np}
\left\{
\begin{array}{rcl}
f_1(m,\mu ,M)&=&\frac
{\displaystyle \left( 1+\mu \right) \left( 2\left( 1-\mu \right)^2 -
\left.  M^2\right/ m^2 \right) }
{\displaystyle 4\zeta\sqrt{\left. M^2\right/ m^2-\left( 1-\mu\right)^2}}
=n
\\
f_2(m,\mu ,M) &=&{\displaystyle \frac
{\displaystyle \left. M^2\right/  m^2-\left( 1-\mu\right)^2}
{\displaystyle 8 \zeta}}-\frac{1}{2}=p
\end{array}
\right.
 \end{equation}
where $n$ is integer and $p$ is positive integer. 
If we take $m_{in}=0$ which corresponds to the Carter-Penrose diagram of Fig.
\ref{fig:spacetime} we obtain the discrete mass spectrum for Schwarzschild mass
$m=m_{out}$ parameterized by two quantum numbers $n$ and $p$ which was found 
in \cite{prd57}.

The first quantum number $n$ comes from ordinary boundary condition on the 
wave function at the infinity in  the classically forbidden region . 
The origin of the second quantum number can be
understood as follows. The classical motion of the shell may be of two types
depending on the values of mass parameters, $m$ and $M$. If $M/m<2$ then we 
have the so called black hole motion. If $M/m>2$, then the shell forms the 
wormhole and it moves beyond the event horizon on the other side of the 
Einstein-Rosen bridge where we have another infinity. 
The configuration 
space in which $R$ takes values is a cross with two ends at the right and 
left space infinities of the Kruscal space-time, two ends in the 
future and past
singularities $R=0$ and an intersection at the horizon $R=2m$.
 But in quantum theory
the wave function covers the regions on both sides of the Einstein-Rosen 
bridge for any values for mass parameters. So we have an additional independent
boundary condition at the second infinity where the classical motion is
also forbidden. It is the boundary condition at the second infinity that 
gives rise to the appearance of the second quantum number.

The quantum number which corresponds to the first quantization
condition disappears for the unbound motion of the shell. 
But the second quantization condition remains valid and we could 
expect that there exist a quantum number for the unbound motion of the shell
as well. Such qualitative considerations are supported by calculations in the 
large black holes \cite{prd57} and quasiclassical \cite{quasi} limits.
 
Analyzing the unbound motion we encounter the following 
difficulty. The solutions of equation (\ref{main}) contain both in-going and 
out-going waves with equal amplitudes. This means that in classical limit we 
have a superposition of space-times with collapsing and expanding shells. But
we'd like to have a single space-time in classical limit. This could be 
obtained by considering the quantum version of equation (\ref{exponent})
  \begin{eqnarray}
\label{nonsymmetric}
\Psi(m, \mu, S+i\zeta )=\nonumber\\
\frac{1}{2\sqrt{F_{in}F_{out}}}
\left( F_{in}+F_{out}-
\frac
{\displaystyle M^2}
{\displaystyle 4 m^2 S}
\pm
\frac
{\displaystyle M}
{\displaystyle m\sqrt{S}}
{\cal Z}
\right)\Psi( m, \mu, S)
 \end{eqnarray}
Analysis analogous to that carried out in \cite{prd57} gives the following 
quantization condition
\begin{equation}
\label{condition}
6f_2+f_1=2\tilde n,\ \tilde n\ -\ integer 
 \end{equation}
where $f_1$ and $f_2$ are functions from (\ref{np}). 
So the quantum number $\tilde n$ is a certain 
combination of quantum numbers for bound motion $n$ and $p$ (\ref{np}).
If we put $m_{in}=0$ we obtain the spectrum for unbound motion which 
corresponds to the situation shown in Fig. \ref{fig:spacetime}b.
The spectrum for unbound motion
depends on the continuous parameter $M$ instead of second discrete quantum number 
as it was explained at the beginning of the paper.

Now let us consider how radiation propagates in the field of black hole. We
want to take into account the effect of back reaction of radiation on
the gravitational. 

We approximate the radiation by a thin null shell. 
The space-time near the light shell is divided
into two regions. The inside region and 
the outside region are both characterized by their Schwarzschild 
masses $m_{in}$ and $m_{out}$. The wave packet takes away a small
amount of the black hole mass $\delta m =m_{out}-m_{in}$. 

The motion of the null shell is described
by the constraint (\ref{exponent}) with mass $M=0$ \cite{nullshell}. 
This means that
 \begin{equation}
\label{light}
\exp\left( \pm G \hat P/R\right) =
\sqrt{F_{in}/F_{out}} 
 \end{equation}
The sign ``$+$'' corresponds to the expanding  motion of the shell and ``$-$''
to the collapsing motion. 

When we go beyond the geometrical optics approximation, we need 
the corresponding wave
equation, which gives trajectories (\ref{light}) in geometric optics limit.
This is just the equation (\ref{nonsymmetric}) with $M=0$. 
 \begin{equation}
\Psi (m, \mu, S-i\zeta )=\sqrt{F_{in}/F_{out}}\Psi(m, \mu, S)
 \end{equation}
The global solutions of this equation exist only if the quantization condition
 \begin{equation}
 \label{sp}
 3\left( 1-\mu\right)^2/8+ \left( 1-\mu^2\right) /4 =\zeta
\tilde n,\ \tilde n \ - integer
 \end{equation}
holds. It is just our Eqn. (\ref{condition}) for $M=0$.
Now if the energy of the shell  
is small $\delta m\ll m_{out}$ , $m_{in}\approx m_{out}\approx m$, 
the Eqn. (\ref{sp}) takes the form
 \begin{equation}
 \label{quantum}
 \delta m \approx m_{pl}^2\tilde n/m
,\ \tilde n - integer
 \end{equation}
This is the radiation spectrum predicted by Bekenstein and Mukhanov \cite{bekenstein}.

The minimal energy of the light shell emitted by a black hole of 
Schwarzschild mass $m$ corresponds to $n=1$:
 \begin{equation}
 \label{min}
\delta m_{min}\approx m_{pl}^2/m\sim T_{BH}
 \end{equation}
It is of order of Hawking temperature $T_{BH}$ for the black hole. 

So we have found that if the process of propagation of radiation emitted
by a black hole is considered in a self-consistent way (the back reaction 
of radiation on gravitational field is taken into account), the spectrum 
of radiation turns out to be discrete and depends only on on the Schwarzschild
mass of black hole (and not on its internal structure). 

The mass energy spectrum of black hole in our model has a rather complicated 
structure ( (\ref{np}) for bound motion and (\ref{condition}) for unbound 
motion). But the spectrum of radiation coming from black hole
is not the spectrum determined by level spacings between mass-energy
levels in the spectrum of black hole. It is because only quanta with energies 
satisfying the requirement (\ref{quantum}) could propagate in the gravitational
field of black hole in a self-consistent way.

Now we come to the problem whether the two spectra are in agreement.
Let a quantum
of energy $\delta m$ (\ref{min}) is emitted by a black hole. Then the 
Schwarzschild mass has decreased: $m'_{out}=m_{out}-\delta m$. So the
parameters $n$, $p$ and $m_{in}$ must tune for the new $m_{out}$. 
This obviously
could not be done only by changing $n\rightarrow n\pm 1$, $p\rightarrow p\pm 
1$. We also have to change $m_{\in}\rightarrow m'_{in}$. So even
if we had $m_{in}=0$ before the emission we will have $m_{in}\not= 0$ 
after the emission. This could have a simple explanation. We have to
conclude that not only a quantum is emitted to the infinity but necessarily
another quantum must be emitted to the inside of the shell. It changes the
mass $m_{in}$ of the shell. 
So the quanta must be born in pairs. The situation is schematically 
presented on Fig. \ref{fig:emission}.

So there are several special features that the  mechanism of
Hawking radiation must possess which could be derived from our model.

1)  Apart from requirement that radiation energy must
correspond to the level differences in mass-energy spectrum (Bohr postulate)
we have to 
consider also the requirement that radiation is able to propagate in the 
gravitational field of the black hole. It is this last requirement that 
leads 
to the fact that a black hole of mass $m$ emits radiation with energies 
of order of Hawking temperature. 
  
2) Quanta of radiation are emitted in pairs. Together with quantum
emitted to the infinity there is always a quantum which falls into the black 
hole. We traced out this second quantum in our model by noting that  the
emission of radiation leads to the change of Schwarzschild mass inside the 
shell. In usual quantum mechanics the process of emission of radiation causes
the transition from one energy level to another in a given quantum-mechanical
system. Here we see that the emission leads to the change in the inner structure 
of black hole and changes the mass spectrum itself.

There is yet another interesting feature of the Hawking radiation in our model.
Let us consider the situation when the mass of black hole is small, that
is at the end of the evaporation process. Then we need the exact formula 
(\ref{sp})
 \begin{equation}
 \left( \delta m\right)^2/8+ m_{out}\delta m/2 =
\tilde n m_{pl}^2/2
,\ \tilde n\ -\ integer
 \end{equation}
This is a quadratic equation for $\delta m$. It's positive solution is
 \begin{equation}
\delta m=-2 m_{out}+ 2\sqrt{m^2_{out}+ \tilde n
m_{pl}^2}
 \end{equation}
The minimal energy of quanta emitted corresponds to $n= 1$. But if
 \begin{equation}
m_{out}<-2 m_{out}+2\sqrt{m^2_{out}+m_{pl}^2}
 \end{equation}
the energy of the emitted quantum would be greater then the mass of the black
hole. So the last feature is

3) Black holes with masses 
 \begin{equation}
m_{out}<\frac{2}{\sqrt{5}} m_{pl}
 \end{equation}
do not radiate. The evaporation of black holes leads in our model 
to  remnants of about Planckian mass.

The main qualitative result of the present letter is the following.
We found that due to the nontrivial structure of configuration space
(which reflects the complex structure of the complete Schwarzschild 
manifold) the selfgravitating null shells have discrete energy spectrum.
We consider such null shell as describing approximately the quantum
Hawking radiation. It appeared that the spectrum of the radiation does
not math the spectrum of the massive selfgravitating thin shell.
The way out of this difficulty is to suppose that collapsing shell 
radiates not only to infinity but also inward. 
the ingoing radiation leads to the change of the mass $m_{in}$ inside
the shell (even it is initially zero). Thus the inner structure of the black
hole is changing during its evaporation. The distant observer however will
see only the radiation spectrum which is that predicted by J.Bekenstein
and V.Mukhanov for black hole masses much larger than the Planckian mass.
As a byproduct we show that the black holes with the masses less than the 
Planckian mass do not radiate.

 \begin{figure}
 \caption{Space-time with a collapsing shell: bound (a) and unbound 
(b) trajectories of the shell.}
 \label{fig:spacetime}
 \end{figure}

 \begin{figure}
 \caption{Process of emission of quanta of radiation.}
 \label{fig:emission}
 \end{figure}

% \begin{table}
% \caption{}
% \label{}
% \end{tabular}
% \end{table}


\begin{references}
\bibitem{kuchar} K.Kuchar, Phys. Rev. {\bf D 50}, 3961, (1994)
\bibitem{prd57} V.A.Berezin, A.M.Boyarsky, A.Yu.Neronov, Phys. Rev. {\bf D} 
57, 1118, (1998) 
\bibitem{landau} L.D.Landau, E.M.Lifshitz, {\it Quantum mechanics}, Nauka,
Moscow, (1974)
\bibitem{quasi} A.Yu.Neronov, {\it Quasiclassical mass spectrum of
quantum black hole model with selfgravitating dust shell.} gr-qc/9808021
\bibitem{nullshell} J.Louko, B.F.Whitting, J.L.Friedman, Phys. Rev. {\bf D 57},
2279, (1998)
\bibitem{bekenstein} J.Bekenstein, A.Mukhanov, Phys.Lett. {\bf B 360}, 7 (1995)
\end{references}
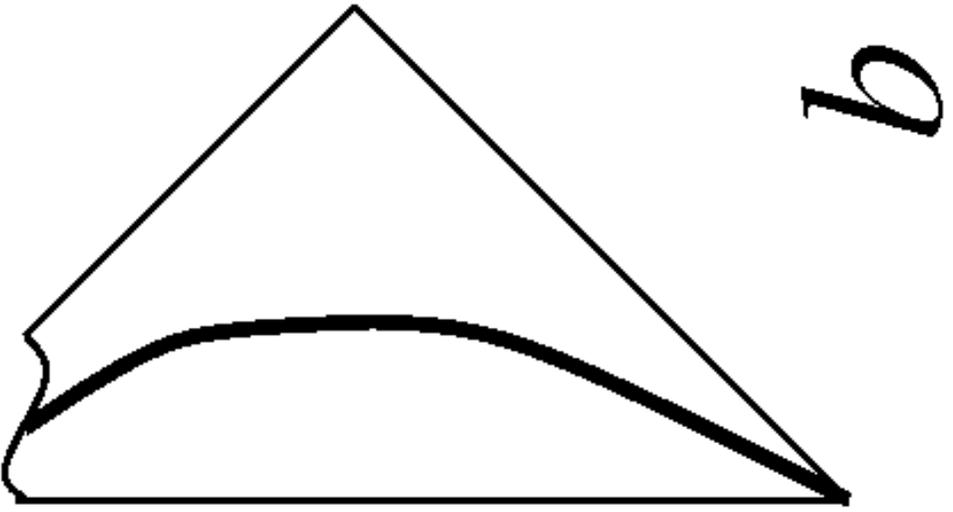
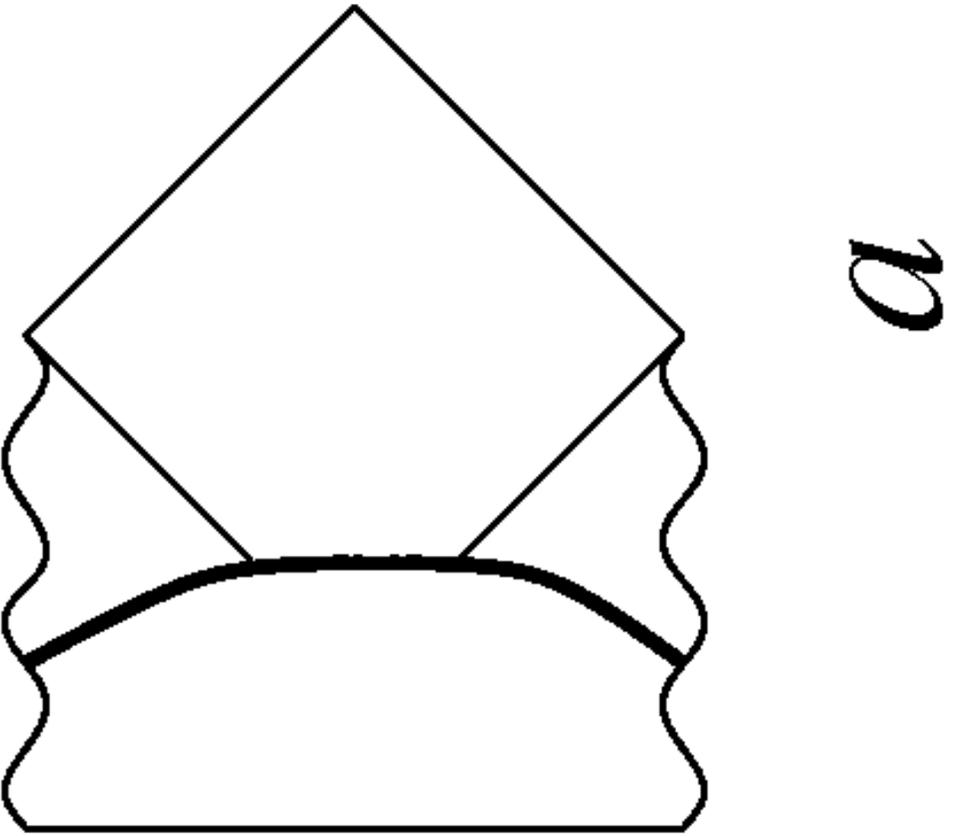
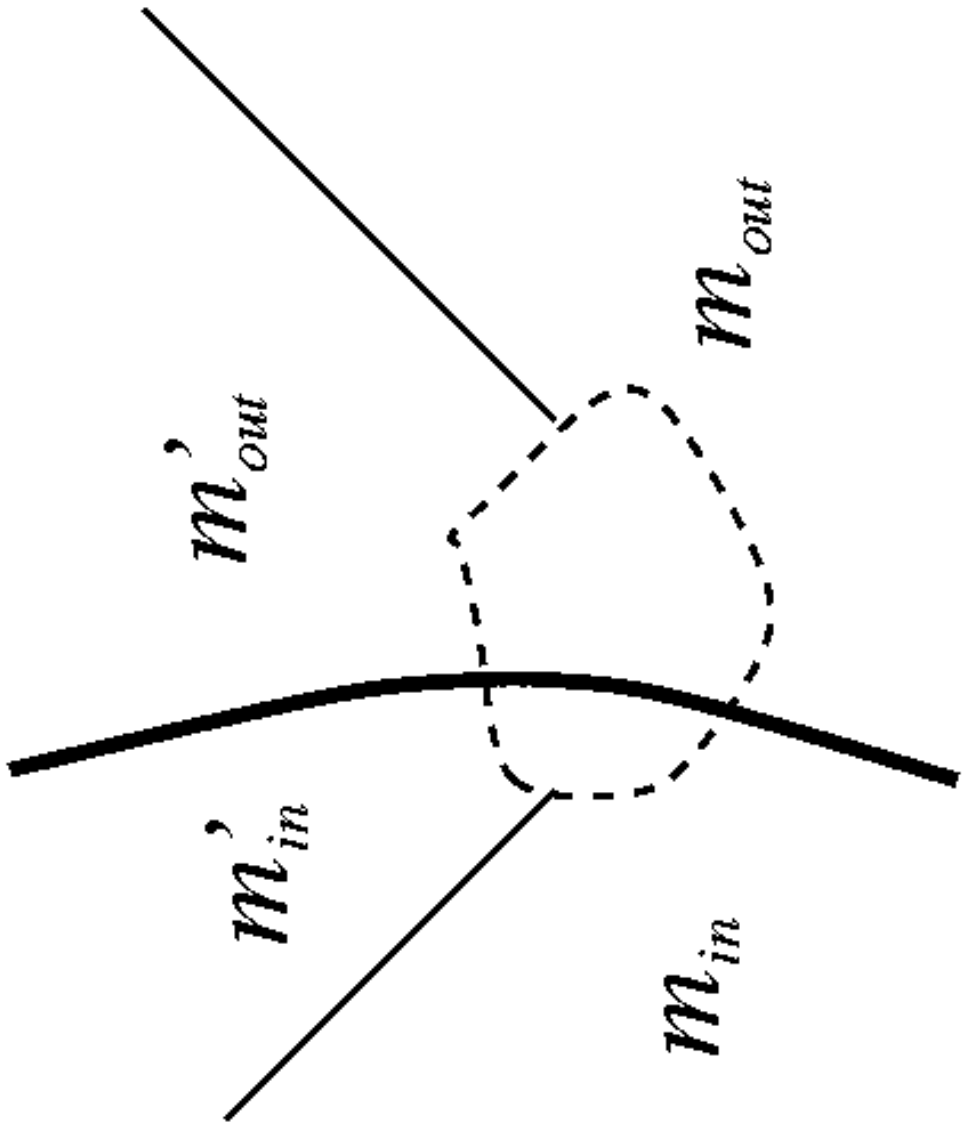
\end{document}